\documentclass[10pt]{article} 
\usepackage{fullpage}

\usepackage{epsfig}
\usepackage{graphicx}
\usepackage{bm}
\usepackage{amsmath}
\usepackage{amsfonts}
\usepackage{amssymb}
\usepackage{latexsym}
\usepackage{psfrag}

\newcommand{\Ket}[1]{\left|#1 \right\rangle}

\begin{document}

\title{\bf Towards quantum computing with single atoms and optical cavities on atom chips}

\author{M. Trupke,$^1$ J. Metz,$^1$ A. Beige,$^2$ and E. A. Hinds$^1$ \\[0.3cm]
{\small $^1$Blackett Laboratory, Imperial College London, Prince Consort Road, London SW7 2BZ, United Kingdom} \\[-0.05cm]
{\small $^2$The School of Physics and Astronomy, University
of Leeds, Leeds LS2 9JT, United Kingdom}}

\date{\today}

\maketitle
\begin{abstract}
We report on recent developments in the integration of optical microresonators into atom chips and
describe some fabrication and implementation challenges. We also review theoretical proposals for
quantum computing with single atoms based on the observation of photons leaking through the cavity
mirrors. The use of measurements to generate entanglement can result in simpler, more robust and
scalable quantum computing architectures. Indeed, we show that quantum computing with atom-cavity
systems is feasible even in the presence of relatively large spontaneous decay rates and finite
photon detector efficiencies.
\end{abstract}

\section{Introduction}

Quantum information processing (QIP) is a new paradigm for
manipulating information. Already, in a first important application
of QIP, quantum cryptography~\cite{bennett,ekert} guarantees the
physically secure transfer of information between distant parties.
As for computing, algorithms have been devised that lead to a
dramatic increase in computational speed when compared to the best
known classical methods \cite{deutsch}. Prominent examples are
Shor's factoring algorithm~\cite{shor} and Grover's database
search~\cite{grover}. If a large-scale quantum computer were to be
realised, with thousands of universal gates, no doubt more
algorithms would emerge. However, even tens of qubits would be
enough to provide a powerful computational platform for simulating
specific quantum systems \cite{simulations} whose complete Hilbert
space is beyond the reach of current digital computers.

Many different physical implementations of QIP are currently being
explored. Over the last few years, several proof-of-principle
experiments have demonstrated the feasibility of quantum computing.
Vandersypen {\em et al.}~\cite{chuang} have realized a simple
instance of Shor's algorithm by factoring $15$ using nuclear
magnetic resonance techniques. Groups in Innsbruck and Boulder have
implemented a universal two-qubit gate in an ion trap and entangled
up to eight ions \cite{blatt,blatt2}. Walther {\em et al.} have
performed linear optics experiments with up to five photonic qubits
and a four-photon cluster state~\cite{walther}. However it is not
straightforward to scale any of these to many more qubits.
Additional qubits in ion traps increase the density of motional
states, thereby creating the need for a form of distributed quantum
computing, possibly involving ion transport~\cite{kielpinski}. As
for linear optics quantum computing, the main difficulties when
entangling photons are the lack of an effective interaction and the
lack of reliable photon storage. For a recent comparative review of
a number of quantum computing implementation proposals
we refer the reader to Meter and Oskin~\cite{meterArchitectures}.

Neutral atoms can be coupled to each other by the quantised field
inside an optical cavity. Although not yet demonstrated
experimentally, this offers a promising alternative implementation
of quantum computing. It has already been shown that single atoms
held in optical resonators are capable of generating single photons
on demand and deterministically, i.e. without spontaneous
emission\cite{kuhn,kimble}. It is therefore possible to exchange
quantum information between a stationary qubit (the atom) and a
flying qubit (the photon), as required by one of DiVincenzo's
criteria for a scalable quantum computing architecture
\cite{DiVincenzo}. This may be used to couple distant parts of a
quantum computer by generating photons whose state depends on the
state of the respective atom \cite{Monroe} followed by carefully
designed photon measurements \cite{grangier,Lim}. 

Photon measurements provide a very efficient tool for manipulating
information in atom-cavity systems. They can be used to simplify
significantly the production of entanglement. One example of that is
the method based on an environment-induced quantum Zeno effect
\cite{letter,Ben,pachos,you}, where the application of laser fields
suffices to entangle two ground-state atoms trapped inside an
optical cavity. A second approach due to Lim {\em et al.} \cite{Lim}
entangles distant atoms by the deterministic generation of photon
pairs and their subsequent detection. As a third example, we
recently we showed that the absence macroscopic fluorescence
\cite{shelving} can signal the presence of maximally entangled atom
pairs \cite{jeremy} and may be used for the successive build-up of
cluster states \cite{new} for one-way quantum computing
\cite{Briegel}. Such a signal can easily be
detected even when using inefficient photon detectors.

Over the last few years, much progress has been made in observing and controlling the electronic
and motional states of atoms inside optical resonators, including a variety of transport
\cite{chapmanTransport, nussmannTransport}, cooling \cite{cooling} and trapping~\cite{kimble,rempe}
mechanisms. For most purposes, the quality of the atom-cavity system is measured by the relative
sizes of the atomic decay rate $\Gamma$, the atom-cavity coupling constant $g$, and the cavity
decay rate $\kappa$. The latter are given by~\cite{kuhnRempe}
\begin{equation}\label{gKappaGamma}
    g=\sqrt{
    \frac{|\mu_{eg}|^2\omega}{2\hbar \epsilon_0 V}} \, , ~~~
    \kappa=\frac{\pi c}{2l F} \, ,
\end{equation}
where $\mu_{eg}$ is the transition dipole moment and $l$ and $F$ are
the length and the finesse of the Fabry-Perot resonator,
respectively. For reflectivities $R > 99\%$, the finesse can be
approximated by $F\approx \pi/(1-R)$. Often, the quality of a
resonator is given using the Q-factor, in terms of which
$\kappa=(\pi c)/(Q \lambda)$. The ratio $C \equiv g^2/\kappa \Gamma$
is known as the single-atom cooperativity parameter. This parameter
is of central importance in atom detection, in single-photon
production, and in the generation of entanglement between atoms
sharing a cavity mode. After several years of steady progress in
atom-cavity experiments, values of $C$ on the order of $50$ have
been achieved \cite{Chapman2,Chapman3}. By measuring a classical
fluorescence signal caused by quantum jumps, as suggested by Metz
{\em et al.} \cite{jeremy}, it should be possible to entangle two
atoms with a fidelity above $90 \%$ even when
$C$ is as low as 10 and when detecting photons with an efficiency as low as $\eta = 0.2$.

\begin{figure}
\begin{center}
{\includegraphics[width=16cm]{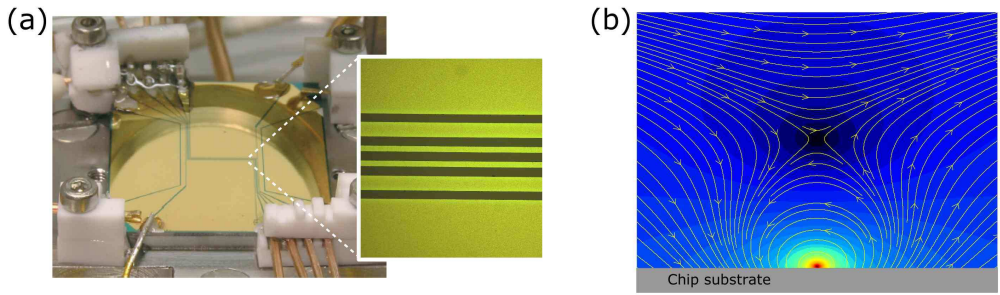}}
\end{center}
\caption{(a) An atom chip used at Imperial College by Eriksson {\em et al.} \cite{ErikssonRev}. (b)
Figure of how single atoms are trapped on an atom chip. The current-carrying wire at the bottom of
the figure (the current is coming out of the plane of the paper) produces a circular magnetic field
which cancels an external field applied from left to right. The resulting field strength is
indicated with light (strong field) and dark (weak field) colours. The atoms are trapped in the
region with the weakest field, in a cylindrical potential parallel to the wire.} \label{chipTrap}
\end{figure}

Atom chips can be manufactured to provide arrays of miniature traps
and guides, allowing for the simultaneous trapping and manipulation
of a large number of spatially separated single atoms
\cite{review_chip}. Fig. \ref{chipTrap} (a) shows an atom chip with
wires fabricated in a gold film. The magnetic fields effected by
current carried by these wires is used to manipulate Bose-Einstein
condensates for atom interferometry \cite{ErikssonRev}. As
demonstrated recently
\cite{aplMicrocav,treutleinChipQIP,aokiMicrotoroid}, optical
cavities that can be integrated into atom chips promise large
cooperativity parameters $C$ due to their small mode volume.
Moreover, neutral atoms are strong candidates for stationary qubit
carriers as they interact only weakly with the environment and
possess long decoherence times. For example, the lifetime of the
coherence between the $(F = 1, m_{F} = -1)$ and $(F = 2, m_{F} = 1)$
hyperfine levels of the $5^2S_{1/2}$ ground state of $^{87}$Rb has
been measured to exceed one second, even when the atoms are held
close to the surface of an atom chip in a magnetic microtrap
\cite{reichelCoherence}. Using silicon as the substrate material
means furthermore that all necessary classical control circuits can
be created on the same chip, leading to a fully integrated device.
Atom-cavity systems on atom chips therefore hold great promise for scalable quantum computing in the near future. 

In this article we report on recent efforts towards quantum
computing with single atoms using optical cavities on atom chips. In
this setup the successful completion of a quantum-logical gate
operation can be heralded either by the deterministic generation and
detection of single photons, the absence of single photon emissions
or the absence of a macroscopic fluorescence signal. Two cavities,
when combined with a reliable transport mechanism such as the
magnetic guides and traps implemented on atom chips, are then
sufficient to carry out quantum-computational operations on a large
number of qubits. In Sections \ref{chip} and \ref{chip2} we give a
brief description of the methods used to trap atoms above an atom
chip and report on recent achievements in combining such systems
with optical resonators. In Sections \ref{sources} and
\ref{entangle} we review recent atom-cavity quantum computing
proposals based on the measurements of photons leaking through the
resonator mirrors. Finally, we summarise our results in Section
\ref{conclusions}.

\section{Guiding and trapping single atoms} \label{chip}

Atom chips are devices with micro-structured surfaces, which produce
magnetic and/or electric fields and enable the trapping, cooling and
manipulation of atomic clouds and single atoms. The magnetic fields
are produced by current-carrying wires or permanently magnetised
surfaces and couple to the magnetic dipole moment of the atoms. The
small scale of the structure produces strong magnetic field
gradients which make tight traps for magnetic atoms
\cite{review_chip}. Figure \ref{chipTrap}(b) shows the field of a
single wire (shown as a dot) to which a uniform bias field has been
added. This creates a zero of the magnetic field above the wire,
surrounded by a region of approximately quadrupole asymmetry. Atoms
in a weak-field-seeking state will be attracted and held in this
region. At a zero of the magnetic field, the weak- and
strong-field-seeking states of the atom are degenerate, so a
transition may occur which would lead to repulsion of atoms from the
trap. To avoid
this, a uniform field can be added parallel to the wire \cite{sinclairVideo,reichelRev}.

Losses can still occur, for example because of current noise caused by thermal fluctuations, though
these can be controlled by a suitable choice of material and film thickness
\cite{JonesHindsSpins,scheel}. The lithographic process involved in creating the wires or
permanent-magnetic structures makes it possible to create complex patterns repeatably and with high
precision, which in turn guarantees the scalability of these components. The wires on atom chips
have typical widths of $1$ to $100\,\mu$m and thicknesses of $1$ to $10\,\mu$m, and can carry
currents on the order of $1$ to $10\,$A. This makes it possible to form strong and tight traps,
with trap frequencies on the order of $10$ kHz and depths on the order of $1\,$mK. Similar trap
characteristics have been obtained with micro-patterned permanent-magnetic surfaces
\cite{sinclairVideo}. Wires patterned using UV-lithography have edges with a feature size of less
than $100\,$nm and with a surface roughness of less than $10\,$nm. This is important to ensure that
electrical currents flow smoothly along the wires, thereby creating a uniform trap for the atoms
\cite{krugerArXiv}. The atoms still need to be pre-cooled before they can be loaded into these
traps, and this is usually done in a magneto-optical trap close to the surface of the chip.

\section{Integrating optical cavities} \label{chip2}

Until recently, atom chip experiments have focussed on trapping and
manipulating large clouds of atoms, with a view to creating
Bose-Einstein condensates (BECs)
\cite{examples,reichelRev,sinclairVideo}. However, the reliable
delivery and individual control of single cold atoms, each of them
carrying one qubit, is a necessity for many quantum computing
schemes. To achieve this, it must first be possible to detect single
atoms with a high degree of confidence \cite{chipSAD}. Furthermore,
as mentioned in the Introduction, the coupling of atoms to the field
of an optical cavity is a powerful tool for entangling them, and the
strength of this coupling increases with decreasing cavity mode
volume. The inherently small mode volume of a microcavity therefore
provides a strong incentive for the use of such a device for quantum
optics experiments. Atom chip circuits enable the positioning of
atoms with high accuracy down to the sub-nanometer scale
\cite{reichelRoySoc}. This is an important tool in the attempt to
couple atoms stably and accurately to the modes of
micro-resonators.

Two types of resonator are being integrated with atom chip technology: whispering-gallery-mode
(WGM) \cite{vernooyKimbleMicrospheres,folmanMicrodisk,reichelRoySoc,Rauschenbeutel,aokiMicrotoroid}
and Fabry-Perot (FP) microcavities \cite{reichelRoySoc, ErikssonRev}. WGM cavities have
unprecedented quality factors, with the best microsphere resonators approaching $Q = 10^{10}$
\cite{vernooyKimbleMicrospheres}. However, because the intense part of the mode is confined within
the solid material of the resonator, coupling to the mode has to be made through the weaker
evanescent field outside. The latter decreases exponentially with distance from the resonator
surface, with a decay constant of order $\lambda/2\pi$. For an atom to interact perceptibly with
the resonator mode it must therefore be placed accurately, i.e.~to within a small fraction of a
wavelength, in close proximity of the resonator surface, where the attractive Van der Waals force
on the atom becomes considerable.

A number of WGM devices have been proposed as candidate systems for the detection and manipulation
of atoms. Fused-silica microspheres have the highest known quality factors, but the procedure used
for fabricating them is not easily included in the production of atom chips. Such a resonator would
have to be positioned on the surface and attached to the atom chip using procedures separate from
the standard etching and coating steps used in the manufacture of semiconductor chip devices.
Furthermore, the dimensions and quality of microspheres vary considerably from one to another
\cite{vernooyKimbleMicrospheres}. Microtoroids are therefore more natural candidates for
integration as they can be produced using standard microfabrication techniques, and still offer
very high quality factors \cite{VahalaMicrocavRev}. Strong coupling between single atoms and the
field of a microtoroid resonator has in fact been demonstrated recently \cite{aokiMicrotoroid},
albeit with atoms passing through the evanescent field in free-fall. While the strong-coupling
condition has been experimentally fulfilled for the first time for a WGM device, the challenge of
reliably positioning atoms in the evanescent field with the required accuracy has yet to be
surmounted.

\begin{figure}
\begin{center}
{\includegraphics[width=16cm]{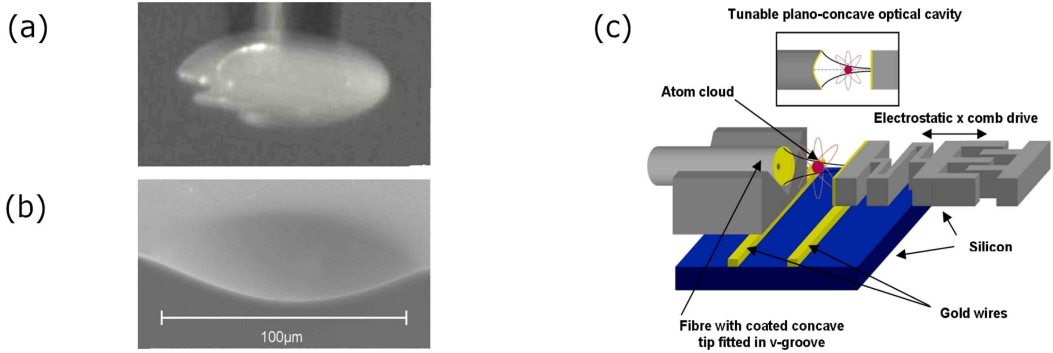}}
\end{center}
\caption{(a) A plane dielectric mirror attached to  the tip of a
single-mode optical fibre. (b) SEM image of a section through the
curved mirror template etched in silicon. (c) schematic of an
integrated tuneable optical microcavity combined with a magnetic
atom trap.} \label{pistol}
\end{figure}

By contrast, FP resonators have lower values of finesse and $Q$. However, atoms can be placed
directly and accurately into the region of highest field strength of the cavity mode, leading in
practice to higher values of $g$ that are reproducible. Furthermore the requirements on positional
accuracy of the atoms within the mode are less stringent because the intensity varies slowly near
the antinodes, with a standing wave spacing of $\lambda /2$ and a mode waist of $3-10 \mu$m. For
these reasons, the efforts of several research groups are currently focussed on this type of
resonator. One type of FP-microcavity, currently in use at Imperial College London, is a
plano-concave resonator consisting of an isotropically etched dip in a silicon surface and the
cleaved tip of a single-mode fibre \cite{aplMicrocav}. Figure \ref{pistol} (a) is a picture taken
under an optical microscope of the coated fibre tip. Initially, the reflecting film is formed by
evaporating onto a donor surface to which it is weakly attached. The fibre tip is then aligned to
this surface, and glued firmly to the reflective coating using a UV-curing, index-matching epoxy.
Pulling the fibre away from the donor surface completes the procedure. This abrupt detachment is
the cause of the rough edges visible in Figure \ref{pistol} (a), but does not damage the reflecting
surface. Figure \ref{pistol} (b) is a scanning electron microscope image of the curved silicon
mirror substrate. The image shows a specimen which has been cleaved close to the centre to make the
curvature more readily discernible. This surface was also subsequently coated with a
high-reflectivity multilayer dielectric film using a standard sputtering procedure. A finesse in
excess of $5000$ and a Q-factor of over $10^6$ have been achieved with cavities of this type. These
values are limited by scattering losses caused by the surface roughness of the silicon mirror
substrate, which is approximately $2\,$nm rms. This can be improved upon by adding a deep reactive
ion etching step, or by depositing and thermally reflowing a layer of silica before applying the
mirror coating. The present performance values are nonetheless sufficient in principle to detect
single atoms with high confidence, and should also enable the generation of single photons with
high efficiency. This method of making concave mirrors relies on standard silicon etching and
coating techniques and is easily included in the chip fabrication procedure. The formation of the
microcavity then only requires the positioning of the coated fibre tip above the chip surface,
without the need for further coupling optics. However, the cavity still needs to be tuned by an
external piezoelectric actuator. The next generation of microcavity will be built on the chip
surface in a planar orientation and will be tuned by an integrated electrostatic
actuator \cite{GollaschJMME2006}.

Several experimental groups have already succeeded in positioning atoms accurately within the mode
of an optical resonator using optical~\cite{kimble,nussmannTransport},
electrostatic~\cite{walthercav,blattcav} and magnetic~\cite{vuleticTransport,haaseTransport}
transport techniques. The positioning of atoms by means of magnetic guides in a microcavity on a
chip has also been recently demonstrated using a fibre-coupled microcavity \cite{treutleinChipQIP}.
Both mirrors of that microcavity are fibre tips to which concave multilayer dielectric coatings
have been applied using a transfer procedure similar to the one described above, but using a convex
donor surface. The highest finesse achieved with this type of microcavity is on the order of
$1000$, limited by mirror roughness. An improved construction uses two fibres with laser-machined
concave tips. They are extremely smooth because the curvature is created by evaporation, which
allows the surface to reflow smoothly as it is formed. This results in a surface roughness of less
than $0.3\,$nm rms. A high-reflection multilayer dielectric coating is then applied to the tips,
giving a finesse of $35000$, and a remarkable theoretical single-atom
cooperativity $C$ of over $250$.

\begin{figure}
\begin{center}
{\includegraphics[width=16cm]{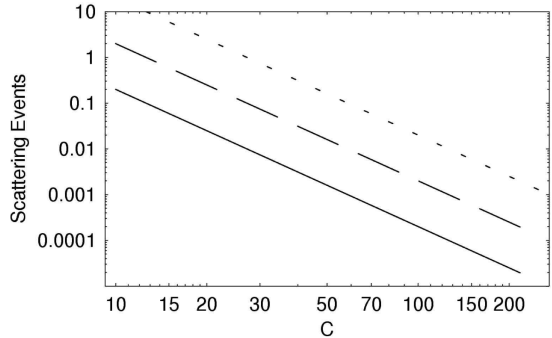}}
\caption{Number of scattering events expected to occur
during the detection of a single atom using a microcavity for a signal-to-noise ratio of 10, and a
detection efficiency of unity (solid line), $10\%$ (long-dashed) and $1\%$ (short-dashed line).}
\label{performance}
\end{center}
\end{figure}

The presence of an atom in the mode of a resonator can drastically
alter its transmission and reflection properties. In microcavities
such as those described above, this effect is large enough to allow
the detection of single atoms with high confidence even for modest
finesse values~\cite{chipSAD, aokiMicrotoroid}. This quality alone
is already of interest for atom-chip experiments as it can be used
to measure {\em in situ} the performance of single-atom transport
and positioning mechanisms available on atom chips. However, beyond
the initial objective of detecting single atoms, it is desirable for
QIP purposes that both the kinetic and internal states of the atom
be preserved beyond the detection event. Microcavities currently
available for atom chips should allow atoms to be detected while
keeping the excitation probability far below unity. For example,
with a weakly pumped system driven on resonance, the number of
scattering events expected to occur during the detection process in
a microcavity with $C\gg1$ is given by \cite{chipSAD}
\begin{equation}\label{scatters}
    M=\frac{S^2}{\eta C^{3}}
\end{equation}
Here $S$ is the signal-to noise ratio and $\eta$ is the detection
efficiency of the system. The latter includes both the quantum
efficiency of the photon detector and losses that occur at any point
after the interaction of the photons with the atom. The solid line
in Figure \ref{performance} shows the number of scattering events
that will occur if the detection efficiency is perfect and the
signal-to-noise ratio is 10. Also shown are curves for 10\% and 1\%
detection efficiencies. In the atom-cavity systems described above,
the primary sources of loss are the roughness of the mirror surfaces
and the mode overlap between the fibre mode and the resonator mode.
Minimising these losses is of the utmost importance not only for the
confident detection of single atoms, but also for the efficiency of
more intriguing applications, such as the deterministic generation
of single photons on demand.

\section{Single photon sources for QIP} \label{sources}

\begin{figure}
\begin{center}
{\includegraphics[width=16cm]{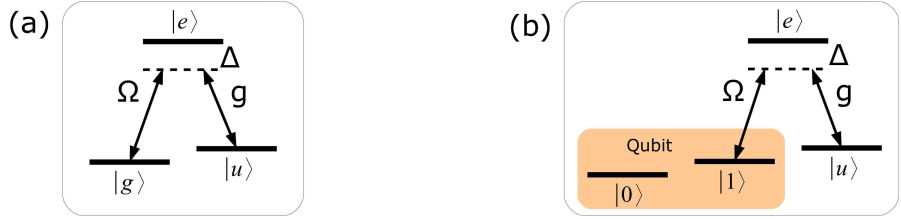}}
\end{center}
\caption{(a) Atomic level configuration for generating a single
photon on demand. The $u$--$e$ transition couples resonantly to the
cavity mode and a laser pulse with an adiabatically increasing Rabi
frequency $\Omega$ drives the $g$--$e$ transition. This transfers an
atom initially prepared in $|g \rangle$ into $|u \rangle$, while
placing exactly one photon into the cavity. (b) An atomic level
configuration, which allows the creation of a photon whose state
({\em early} or {\em late}) encodes the state of the qubit contained
in the atomic ground states $|0 \rangle$ and $|1 \rangle$.}
\label{split2}
\end{figure}

Atom-cavity systems such as those presented above have sufficiently
high cooperativity to generate single photons on
demand\cite{kuhn,kimble}. For this to occur deterministically, it
must be possible to position a single atom at a maximum of the
cavity mode. The relevant atomic level configuration is shown in
Figure \ref{split2}(a). The system is initially prepared with the
atom in state $|g \rangle$ and no light in the cavity. A cavity
photon is then generated by applying a laser beam with increasing
Rabi frequency $\Omega$, as shown in Fig.~\ref{split2}(a), in order
to drive a stimulated Raman transition~\cite{kuhn}. This takes the
atom adiabatically from $|g \rangle$ to $|u \rangle$ without
creating any significant population of the excited state $|e
\rangle$, thereby coherently scattering a single photon from the
laser field into the cavity mode. This photon subsequently leaks out
of the cavity through the
output coupling mirror.

Single-photon sources have many potential applications in QIP. One
is the generation of entangled multi-photon states \cite{jmo} for
linear optics quantum computing \cite{KLM}, where each qubit is
encoded in a photon and gate operations are effected by
measurements. Photonic qubits are attractive because they are easily
distributed and all the required operations for quantum computing
are based on existing technology. Currently, linear optics
experiments rely on photon pairs produced by parametric
down-conversion. Detection of one photon heralds the presence of the
other with certainty, but still the arrival times are uncontrolled,
making it necessary to store photons for substantial periods of time
in order to build up states of several entangled photons. The
current maximum number is five~\cite{walther}.

Beyond its use as a single-photon source, the atom-cavity system can
form a link between stationary and flying qubits. If information is
encoded in the internal states of the atomic sources
\cite{cirac,grangier}, two distant qubits can be coupled by photons
generated in their respective cavities. A carefully designed
photon-pair measurement, i.e. one which does not reveal any
information about the individual sources, can then entangle the
qubits {\em deterministically}, as proposed by Lim {\em et al.}
~\cite{Lim}. Quantum computing using this scheme is scalable, even
with only two cavities.  It is possible to perform a universal
two-qubit gate operation between any two atoms above an atom chip
provided any
atom can be moved at will into one of the resonators. 

One atomic level configuration suitable for implementing
this~\cite{Lim} is shown in Figure~\ref{split2}(b). Each qubit is
encoded in two stable ground states, labelled $|0 \rangle$ and $|1
\rangle$, of a single atom trapped inside an optical cavity. A
comparison with Figure \ref{split2}(a) shows that this level scheme
can be operated as a source for the generation of single photons on
demand. A photon can be created if the atom is initially prepared in
$|1 \rangle$, while the system cannot generate a photon if the state
of the atom is $|0 \rangle$. A laser pulse first swaps the states
$|0 \rangle$ and $|1 \rangle$, followed by an increasing laser pulse
for the generation of a single photon on demand. If this process is
repeated, a qubit initially prepared in $\alpha \, |0 \rangle +
\beta \, |1 \rangle$ becomes
\begin{eqnarray} \label{enc}
\alpha \, |0 \rangle + \beta \, |1 \rangle &
\longrightarrow & \alpha \, |0, {\sf E} \rangle + \beta \,
|1, {\sf L} \rangle \, ,
\end{eqnarray}
where $|{\sf E} \rangle$ and $|{\sf L} \rangle$ denote an {\em early} and a {\em late} generated
photon, respectively. This encoding step (\ref{enc}) entangles the qubit with a newly generated
photon. A measurement on the photon therefore also affects the state of the atomic qubit. 

If two atoms are initially prepared in an arbitrary two-qubit state of the form $\alpha \, |00
\rangle + \beta \, |01 \rangle + \gamma \, |10 \rangle + \delta \, |11 \rangle$, then the encoding
step (\ref{enc}) transforms the system according to
\begin{eqnarray}
\alpha \, |00 \rangle + \beta \, |01 \rangle  + \gamma \,
|10 \rangle + \delta \, |11 \rangle & \longrightarrow &
\alpha \, |00, {\sf EE} \rangle + \beta \, |01, {\sf EL}
\rangle  + \gamma \, |10, {\sf LE} \rangle + \delta \, |11,
{\sf LL} \rangle \, .
\end{eqnarray}
A suitably designed measurement then projects the photon pair into a
state of the form $|{\sf EE} \rangle + {\rm e}^{{\rm i} \varphi_1}
\, |{\sf EL} \rangle  + {\rm e}^{{\rm i} \varphi_2} \, |{\sf LE}
\rangle + {\rm e}^{{\rm i} \varphi_3} \, |{\sf LL} \rangle$, with
the result
\begin{eqnarray}
\alpha \, |00 \rangle + \beta \, |01 \rangle  + \gamma \,
|10 \rangle + \delta \, |11 \rangle & \longrightarrow &
\alpha \, |00 \rangle + \beta \, {\rm e}^{- {\rm i}
\varphi_1} \, |01 \rangle  + \gamma \, {\rm e}^{ - {\rm i}
\varphi_2} \, |10 \rangle + \delta \, {\rm e}^{- {\rm i}
\varphi_3} \, |11 \rangle \, .
\end{eqnarray}
This final state differs from the initial state by a unitary phase
gate. Performing a phase gate in a deterministic fashion therefore
requires only photon-pair measurements, where each of the four
possible measurement outcomes is an equal superposition of the
states $|{\sf EE} \rangle$, $|{\sf EL} \rangle$, $|{\sf LE} \rangle$
and $|{\sf LL} \rangle$. With linear optical elements alone, it is
possible to perform these measurements on a basis of two maximally
entangled states and two product states\cite{Lim}. The detection of
a maximally entangled photon state indicates the realisation of an
entangling two-qubit phase gate on the atoms. This is equivalent to
a controlled-Z gate up to local phase shifts. The detection of a
product state, on the other hand, indicates the realisation of a
local phase gate. Since the atomic qubits are not destroyed at any
stage of the computation, the implementation
of a desired universal entangling phase gate can be repeated until success. 

Under realistic conditions, repeat-until-success quantum gates are
less than $100\%$ successful because of photon loss. Nevertheless,
the scheme of \cite{Lim} can be used to build up cluster
states~\cite{sean,minks} with very high fidelity. The detection of a
photon pair perfectly heralds the outcome of the gate operation on
the atoms as long as dark counts are negligible. Two-dimensional
cluster states constitute a very efficient resource for quantum
computing. Once a cluster state has been built, a whole quantum
computation can be performed using only single-qubit rotations and
single-qubit measurements \cite{Briegel}.

\section{Atom-cavity schemes for QIP} \label{entangle}
Instead of entangling atoms in separate cavities, one could achieve
entanglement with the atoms placed at two antinodes within the same
cavity. This can be done coherently but dissipative processes, which
are unavoidable in real resonators, are detrimental. As Zheng and
Guo \cite{zheng} have pointed out, when one tries to minimise
spontaneous emission with the help of large detunings, this comes at
the expense of slower gate operations. In the end the failure rate
is independent of detuning and depends primarily on the single-atom
cooperativity parameter $C$. The scheme was nonetheless successfully
implemented with Rydberg atoms flying through a high finesse cavity \cite{haroche}.

In order to lower the requirements on the cooperativity of
atom-cavity systems, dissipation can be employed {\em
constructively} for quantum gate operations by performing
appropriate measurements on the system. The first example was the
1995 atom-cavity quantum computing scheme of Pellizzari {\em et
al.}~\cite{Pellizzari}, based on a dissipation-assisted adiabatic
passage \cite{Marr}. A hybrid approach using dissipation in the form
of an environment-induced quantum Zeno effect, whereby the evolution
of the system is inhibited by frequent measurement, and adiabatic
passages was suggested by Pachos and Walther \cite{pachos}. They
predict gate success rates above $85 \%$ even for $C = 100$. This
improvement comes at the expense of a relatively complex stimulated
Raman adiabatic passage (STIRAP) entangling process. A related but
simpler scheme by Yi {\em et al.} \cite{you} achieves gate success
rates above $80 \%$ for $C = 250$. Most recently we have shown (see
below) that even better performance can be achieved if the atoms are
coupled resonantly to the
cavity mode. 

In the following, we summarise the scheme by Yi {\em et al.}
\cite{you}, which achieves a controlled phase gate between two atoms
trapped inside an optical cavity without having to address the atoms
individually. A slight modification of the scheme \cite{you} allows
us to achieve gate success rates above $90 \%$ even for $C = 250$.
The scheme is based on a quantum Zeno effect combining ideas in
Refs.~\cite{letter,pachos} and \cite{you}. While the scheme by Lim
{\em et al.} \cite{Lim}, discussed in Section \ref{sources}, uses
photon detections to impose an entangling gate operation, the same
goal is now achieved by observing the absence of emissions. For $C
\gg 100$, the quantum Zeno effect significantly reduces the
probability for an emission to take place and schemes based on this
effect therefore do not depend on having exceptionally efficient
single-photon detectors. Nevertheless, the proposed scheme makes use
of measurements to simplify
the realisation of the gate. 

\begin{figure}
\begin{center}
{\includegraphics[width=12cm]{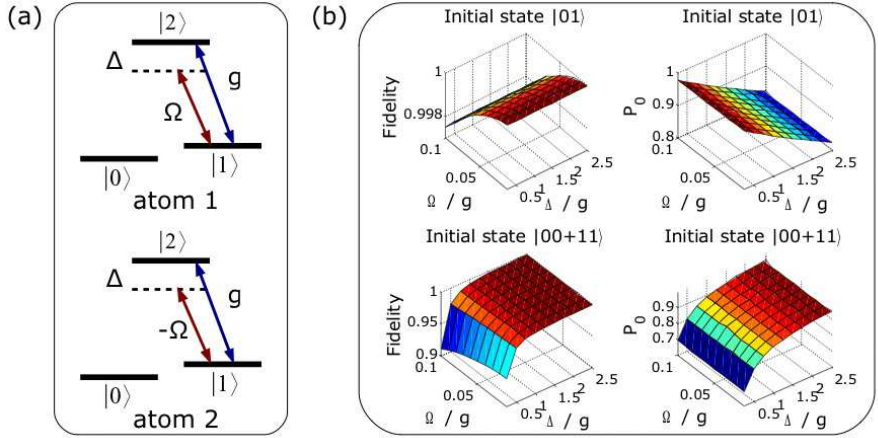}}
\end{center}
\caption{(a) Level configuration for realising a controlled phase gate between two atoms trapped
inside an optical cavity without individual laser addressing. Each qubit is obtained from two
different ground states of an atom. A laser field with detuning $\Delta $ excites the 1--2
transition of each atom, which is in resonance with the cavity field. Atom 1 sees the Rabi
frequency $\Omega$, while atom 2 experiences the Rabi frequency $- \Omega$. (b) Performance
analysis of a single phase gate for the initial states $|01 \rangle$ and $( |00\rangle + |11
\rangle )/\sqrt{2}$. The left hand side shows the fidelity under the condition of {\em no} photon
emission, while the right hand side shows the gate success rate as a function of the Rabi frequency
$\Omega/g$ and detuning $\Delta/g$, and for the same parameters as in Ref.~\cite{you}, namely
$\kappa = 0.05 \, g$ and $\Gamma = 0.08 \, g$, i.e.~$C = 250$.} \label{zeno}
\end{figure}

Suppose two atoms with a $\Lambda$-like level configuration as shown
in Figure \ref{zeno}(a) are simultaneously trapped inside an optical
cavity. The 1--2 transition of each atom couples with the same
coupling constant $g$ to the field mode inside the resonator. Then
there exists a five-dimensional subspace whose population cannot
emit a photon into the cavity. This subspace of so-called dark
states is spanned by the qubit states $|00 \rangle$, $|01 \rangle$,
$|10 \rangle$, $|11 \rangle$ and the maximally entangled
antisymmetric state
\begin{eqnarray}
\Ket{a_{12}} &\equiv & \big( \Ket{12}-\Ket{21} \big)/\sqrt{2} \, ,
\end{eqnarray}
which does not couple to the cavity mode due to the symmetry of the state \cite{letter}. When
prepared in $|a_{12} \rangle$, both atoms try to place excitation into the cavity field but
interfere destructively so that neither of them succeeds. 

If the atom-cavity coupling $g$ and the cavity decay rate $\kappa$ are strong enough, then the
observation of no photons leaking out through the cavity mirrors constitutes a continuous
measurement of whether the system is in a dark state or not \cite{letter,Ben}. The application of a
weak perturbation $H_{\rm int}$ under these conditions cannot move the system out of the
decoherence-free subspace because the system evolves according to the effective Hamiltonian
\begin{eqnarray} \label{He}
H_{\rm eff} &=& I\!\! P_{\rm DS} \, H_{\rm int} \, I\!\! P_{\rm DS} \, ,
\end{eqnarray}
where $I\!\! P_{\rm DS}$ is the projector onto the dark states of the system,
\begin{eqnarray} \label{PDS}
I\!\! P_{\rm DS} &\equiv & |00 \rangle \langle 00| + |10 \rangle \langle 10| + |01 \rangle \langle
01| + |11 \rangle \langle 11| + |a_{12} \rangle \langle a_{12}| \, .
\end{eqnarray}
To obtain the interaction $H_{\rm int}$, laser fields in any
configuration could be applied, as long as their Rabi frequencies
are sufficiently weak compared to $g$ and $\kappa$. The effective
Hamiltonian (\ref{He}) realises a quantum gate between the qubits if
the timing is such that there is no population in $|a_{12} \rangle$
at the end of the operation. One could for example implement a
CNOT-gate by driving the 1--2 transition in atom 2 with a Rabi
frequency $\Omega_1$ and the 0--2 transition in atom 1 with a Rabi
frequency $\Omega_0$ \cite{Ben}. However, it is also possible to
perform two-qubit phase gates without addressing the atoms
separately, as we now discuss. Furthermore, we can stay in the same
parameter regime as in Ref.~\cite{pachos},
\begin{equation} \label{par}
g^2 / \kappa \, , ~ \kappa > |\Delta | \gg |\Omega | \, ,  ~ \Gamma
\, .
\end{equation}
In the following, a laser field drives the 1--2 transition in atom 1
with Rabi frequency $\Omega$, while atom 2 is driven with the
opposite phase to make a Rabi frequency of $-\Omega$, as illustrated
in ~Figure \ref{zeno}(a). The detuning $\Delta $ can be used to
control the population of excited state $|a_{12} \rangle$ and thus
to ensure low spontaneous emission out of the cavity, but there is
no need to use it to avoid populating the cavity mode. In an
appropriately chosen interaction picture, the interaction
Hamiltonian $H_{\rm int}$ can then be written as
\begin{eqnarray} H_{\rm int} &=&
{\textstyle {1 \over 2}} \hbar \Omega \, \big( \, |1 \rangle_{11} \langle 2| - |1 \rangle_{22}
\langle 2| + {\rm H.c.} \, \big) + \hbar \Delta \, \sum_{i=1}^2  |2 \rangle_{ii} \langle 2| \, .
\end{eqnarray}
Together with Eqs.~(\ref{He}) and (\ref{PDS}), this yields the effective Hamiltonian
\begin{eqnarray} \label{eff}
H_{\rm eff} &=& {\textstyle {1 \over 2}} \hbar \sqrt{2} \Omega \, \big( \, |11 \rangle \langle
a_{12}| + {\rm H.c.} \, \big) + \hbar \Delta \,  |a_{12} \rangle \langle a_{12}| \, .
\end{eqnarray}
Given the parameter regime (\ref{par}), the Hamiltonian (\ref{eff}) can be simplified further via
an adiabatic elimination of the excited atomic state $|a_{12} \rangle$. This yields
\begin{eqnarray}
H_{\rm eff} &=& \hbar \Delta_{\rm eff} \, |11 \rangle \langle 11| \, ,
\end{eqnarray}
where $\Delta_{\rm eff} \equiv - \Omega^2 / (2 \Delta)$. The corresponding time evolution operator
equals
\begin{eqnarray}
U_{\rm eff}(T,0) &=& |00 \rangle \langle 00| + |01 \rangle \langle 01| + |10 \rangle \langle 10| +
{\rm e}^{{\rm i} \Delta_{\rm eff} T} |11 \rangle \langle 10|
\end{eqnarray}
and adds a minus sign to the state $|11 \rangle$, if $T = \pi/|\Delta_{\rm eff} |$. A single laser
pulse can therefore indeed be used to realise a controlled phase gate with a very high fidelity
even in the presence of non-negligible spontaneous decay rates $\kappa$ and
$\Gamma$.

This two-qubit gate has the advantage of being highly successful
even for relatively moderate cavity parameters. Figure \ref{zeno}(b)
shows the fidelity and success rate versus $\Delta /g$ and $\Omega
/g$ of a controlled phase gate with $C = 250$ for the initial states
$|01 \rangle$ and $(|00 \rangle + |11 \rangle)/\sqrt{2}$. In
particular, we see that the system is optimally operated using
$\Delta \approx 1.25 \, g$, thereby maximising both the fidelity and
the no-photon probability. Using ideal photon detectors, it is
possible to achieve gate fidelities well above $99 \%$ under the
condition of no photon emission, and gate success probabilities
above $90 \%$. If the detectors are less efficient the gate
fidelity decreases, but only to the level of the gate success rate. 

\begin{figure}
\begin{minipage}{\columnwidth}
\begin{center}
{\includegraphics[width=12cm]{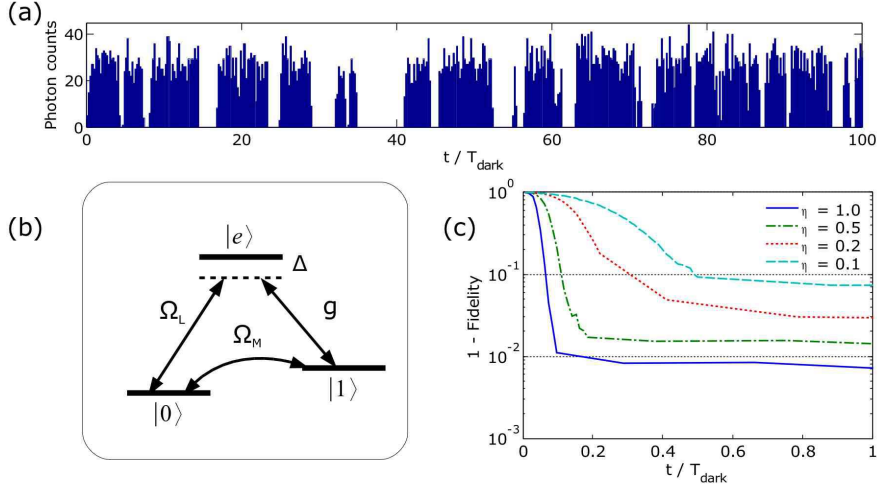}}
\end{center}
\caption{(a) Macroscopic fluorescence signal produced in a simulation with $C = 40$. During dark periods, the atoms are in the maximally entangled state (\ref{a}). (b) Level configuration of a single atom. Both atoms see the same laser Rabi frequencies $\Omega_{\rm M}$ and $\Omega_{\rm L}$ and experience the same atom-cavity coupling constant $g$ and detuning $\Delta$. (c) Fidelity of the final state prepared upon the detection of no photon for a time $t$ for various detector efficiencies $\eta$.}
\label{split3}
\end{minipage}
\end{figure}

Recently, Metz {\em et al.} \cite{jeremy} proposed an alternative measurement-based scheme for the
generation of a well-defined maximally entangled state of two atoms. Instead of deciding on the
success of the operation by the absence or the detection of single photons, the scheme uses
macroscopic light and dark periods \cite{shelving}, during which the system either emits no photons
at all (dark period) or generates photons stochastically (light period). In the following, we
consider a system designed such that macroscopic dark periods, like the ones in Figure
\ref{split3}(a), indicate the successful preparation of two atoms in a maximally entangled atomic
ground state. Since a macroscopic dark period can easily be distinguished from a macroscopic light
period, the scheme is expected to work reliably even when using imperfect photon detectors. 

The telegraph signal in the fluorescence of a single quantum system
was discussed as early as 1975 by Dehmelt in the context of electron
shelving \cite{shelving}. At that time, the origin of the underlying
process raised many questions about our understanding of quantum
mechanics. Nowadays it is well known that the interaction of the
system with its environment effectively results in a measurement,
thereby revealing information about the system. This projects it
into a state which is characterised either by constant
fluorescence or by its complete absence \cite{Cook,Hegerfeldt}. 

Figure \ref{split3}(b) shows the level configuration of a single
atom. A system composed of two such atoms in a cavity is expected to
exhibit macroscopic light and dark periods. Again, each qubit is
obtained from the ground states $|0 \rangle$ and $|1 \rangle$ of an
atom. The 1--$e$ transition couples with detuning $\Delta$ to the
cavity mode, while the 0--$e$ transition and the 0--1 transition are
driven by fields with Rabi frequencies $\Omega_{\rm L}$ and
$\Omega_{\rm M}$, respectively. Given the parameter regime
$|\Omega_{\rm M}| < g , \, \kappa , \, \Gamma  , \, |\Omega_{\rm L}|
\ll \Delta$ \cite{jeremy}, the system possesses the dark state
\begin{eqnarray} \label{a}
\Ket{a_{01}} &\equiv & \big( \Ket{01}-\Ket{10}
\big)/\sqrt{2} \, .
\end{eqnarray}
In the presence of the driving fields, this state is the only one which does not emit photons into
the cavity mode. The observation of a macroscopic dark period in the fluorescence of photons
through the cavity mirrors therefore reveals the successful preparation of the atoms in (\ref{a}).
Switching off the driving fields is then sufficient to keep the system in this maximally entangled
state. 

Below we list the typical time scales considered, with $T_{\rm cav}$ denoting the mean time between
photons during a light period, while $T_{\rm light}$ and $T_{\rm dark}$ are the mean durations of
light and dark periods, respectively. As shown in Ref.~\cite{jeremy}, they are given by
\begin{eqnarray} \label{times}
T_{\rm cav}  = {3 \kappa \Delta^2 \over 4 g^2 \Omega_{\rm L}^2}  \, , ~~~ T_{\rm dark}= {64 \over 9} \, C \cdot
T_{\rm cav} ~~~ {\rm and} ~~~ T_{\rm light} = {64 \over 3} \, C
\cdot T_{\rm cav}  \, ,
\end{eqnarray}
if $\Omega_{\rm L}^2/4 \Delta \Omega_{\rm M} \ll 1$. Here we have
assumed for simplicity that the 0--$e$ and the 1--$e$ transition
have the same spontaneous decay rates. The ratio of $T_{\rm dark}$
to $T_{\rm cav}$ is crucial for distinguishing a light from a dark
period. Eq.~(\ref{times}) shows that it is possible to have $T_{\rm
dark}$ almost 300 times as long as $T_{\rm cav}$, even when the
single-atom cooperativity parameter is as low as 40. Since $T_{\rm
dark}$ and $T_{\rm light}$ are of about the same size, one does not
have to wait very long before the system assumes the desired state
(\ref{a}). 

Figure \ref{split3}(c) shows the fidelity of the state that is
prepared if the laser field is turned off upon the detection of no
photon for a time $t$. Fidelities above $95 \%$ are achievable even
with a detector efficiency as low as $10 \%$ and for the realistic
cooperativity $C = 40$. This method is robust because it makes use
of the dark periods of the telegraph signal. Instead of being a
destructive effect, dissipation plays a key role in the protocol to
generate entanglement. Work is currently underway to use macroscopic
light and dark periods for the generation of cluster states for
one-way quantum computing \cite{new}.

\section{Conclusions} \label{conclusions}

In Sections \ref{chip} and \ref{chip2}, we have reported on recent
developments in atom chip technology and have described the
integration of optical cavities into atom chips. Microcavities
currently available for this purpose have already successfully
detected single atoms, and are of sufficient quality to be used for
the generation of single photons on demand. Sections \ref{sources}
and \ref{entangle} outlined possible QIP experiments with single
atoms and pairs of atoms trapped in the field of optical cavities.
In all examples, atoms are entangled by measuring photons leaking
through the resonator mirrors. The successful completion of an
operation can be heralded by the detection of single photons, the
absence of photon emissions or the absence of a macroscopic
fluorescence signal. Using measurements to herald entanglement leads
to a considerable easing of experimental constraints and increases
the robustness and scalability of
quantum computing architectures. \\[0.5cm]

\noindent {\em Acknowledgement.} A. B. acknowledges funding from the Royal Society and the GCHQ as
a James Ellis University Research Fellow. This work was supported in part by the UK Engineering and
Physical Sciences Research Council through its interdisciplinary Research Collaboration on Quantum
Information Processing, and by the European Union Network SCALA.

\end{document}